\documentclass{vgtc}                          

\graphicspath{{figures/}{pictures/}{images/}{./}} 

\usepackage{times}                     

\usepackage{tabu}                      
\usepackage{booktabs}                  
\usepackage{lipsum}                    
\usepackage{mwe}                       
\usepackage{multirow} 
\usepackage{amsmath}
\usepackage{mathptmx} 

\usepackage{mathptmx}                  

\onlineid{0}

\vgtccategory{Research}

\vgtcinsertpkg

\title{Mazed and Confused: A Dataset of Cybersickness, Working Memory, Mental Load, Physical Load, and Attention During a Real Walking Task in VR}

\author{Jyotirmay Nag Setu\thanks{e-mail: jyotirmaynag.setu@utsa.edu}\\ %
        \scriptsize The University of Texas at San Antonio %
\and Joshua M Le\thanks{e-mail:  joshua.le@my.utsa.edu}\\ %
     \scriptsize The University of Texas at San Antonio %
\and Ripan Kumar Kundu\thanks{e-mail: rkcgc@missouri.edu}\\ %
     \scriptsize University of Missouri-Columbia %
\and Barry Giesbrecht \thanks{e-mail: giesbrecht@ucsb.edu}\\ %
     \scriptsize University of California, Santa Barbara %
\and Tobias Höllerer \thanks{e-mail: holl@cs.ucsb.edu}\\ %
     \scriptsize University of California, Santa Barbara %
\and Khaza Anuarul Hoque \thanks{e-mail: hoquek@missouri.edu}\\ %
     \scriptsize University of Missouri, Columbia %
\and Kevin Desai \thanks{e-mail: kevin.desai@utsa.edu}\\ %
     \scriptsize The University of Texas at San Antonio %
\and John Quarles \thanks{e-mail: John.Quarles@utsa.edu}\\ %
     \scriptsize The University of Texas at San Antonio %
     }

\teaser{
  \centering
  \includegraphics[width=\linewidth, alt={An Overview of the Maze Study}]{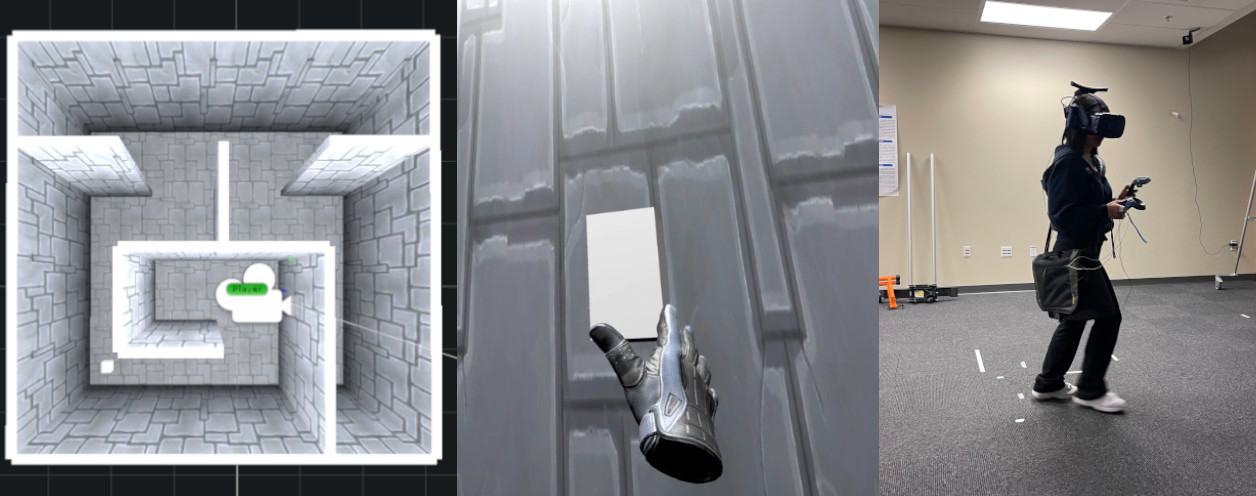}
  \caption{%
  	a) Top-down view of a randomly generated virtual maze environment b) First-person point of view - a participant grabs the trophy to start another random maze task c) Participants physically walked in the real environment to enable  virtual environment navigation%
  }
  \label{fig:teaser}
}

\abstract{
Virtual Reality (VR) is quickly establishing itself in various industries, including training, education, medicine, and entertainment, in which users are frequently required to carry out multiple complex cognitive and physical activities. 
However, the relationship between cognitive activities, physical activities, and familiar feelings of cybersickness is not well understood and thus can be unpredictable for developers. Researchers have previously provided labeled datasets for predicting cybersickness while users are stationary, but there have been few labeled datasets on cybersickness while users are physically walking. Moreover, it is unclear how walking while cybersick will affect cognitive load, even though room-scale interaction is typical in many VR games. Thus, from 39 participants, we collected head orientation, head position, eye tracking, images, physiological readings from external sensors, and the self-reported cybersickness severity,  physical load,  and mental load in VR. Throughout the data collection, participants navigated mazes via real walking and performed tasks challenging their attention and working memory. To demonstrate the dataset's utility, we conducted a case study of training classifiers in which we achieved 95\% accuracy for cybersickness severity classification. The noteworthy performance of the straightforward classifiers makes this dataset ideal for future researchers to develop cybersickness detection and reduction models. To better understand the features that helped with classification, we performed SHAP(SHapley Additive exPlanations) analysis, highlighting the importance of eye tracking and physiological measures for cybersickness prediction while walking. This open dataset can allow future researchers to study the connection between cybersickness and cognitive loads and develop prediction models. This dataset will empower future VR developers to design efficient and effective Virtual Environments by improving cognitive load management and minimizing cybersickness.
} 


\begin{document}


\firstsection{Introduction}

\maketitle

\par Human-computer interaction now has new avenues to explore, thanks to the development of immersive virtual reality technology. From education, healthcare, and industrial training to gaming\cite{checa2020review,radianti2020systematic,javaid2020virtual}, these technologies can potentially change several industries. However, individuals may encounter physiological and psychological difficulties when they immerse themselves in these virtual experiences. In many popular VR experiences, players are required to perform multiple tasks concurrently, some of which are physical and some of which are mental \cite{armougum2019virtual, vzagar2020human}. For example, consider the popular room-scale VR game Half Life: Alyx, in which players must battle enemies (physical), solve puzzles (mental), and navigate unfamiliar environments (physical and mental). For a developer, finding a balance between mental and physical demands in VR can be a daunting task that currently requires extensive playtesting. 
Unfortunately, there are few comprehensive datasets in the literature that can be used to help predict and combat physical and mental demands on VR users. To address this research gap, this paper provides a novel dataset intended to enable the assessment and prediction of cybersickness, working memory, mental load, physical load, and attention during room-scale VR experiences.

Few prior works have investigated the intersection of cybersickness, cognition, and real walking\cite{varmaghani2022spatial,kim2023exploring}. For example, several cybersickness-related works have focused on collecting datasets for cybersickness prediction in VR  \cite{islam2020automatic,islam2021cybersickness} while users were mostly stationary and did not consider the impact on cognition. Luong et al.\cite{luong2022demographic} ran a large lab-in-the-field study with participants who used real-walking to navigate a 20x20 ballroom, but they failed to address the potential impact of cybersickness on cognition. In contrast to the prior work, our approach was to collect cybersickness and cognitive data while participants were physically walking. 

Real walking in VR is becoming more commonplace due to off-the-shelf hardware support for room-scale locomotion, wide area navigation, and location-based VR, which can utilize a large amount of physical space\cite{sayyad2020walking}. Suma et al. \cite{suma2009evaluation} pointed out that in scenarios demanding swift and efficient navigation or travel mirroring real-world movements, genuine walking offers benefits over conventional joystick-driven virtual traversal methods. In addition to this, Suma et al. demonstrated dual tasks could serve as valuable indicators for assessing the impact of VR and cybersickness on working memory, attention, and cognitive load. Although many studies have shown that cybersickness is often reported as lower during real walking tasks\cite{wang2021development} compared to stationary tasks, cybersickness can still be present in real walking tasks\cite{lohman2022evaluating}. Thus, cybersickness during real walking tasks may still affect user experience, task performance, and potentially cognitive load.


As a case study to demonstrate the utility of the dataset,  we have evaluated the accuracy of several deep-learning models in classifying cybersickness as most of the currently available datasets are focused on multimodal cybersickness prediction\cite{islam2020automatic}. Additionally, we have conducted SHAP (SHapley Additive exPlanations) explainable-AI analysis to identify the primary features utilized by the deep learning models. In short, our contribution includes:
\begin{itemize}
    \item  VRWalking - A novel open dataset including VR images, eye-tracking, head-tracking, Heart Rate(HR), Galvanic Skin Response(GSR), cybersickness, mental load, physical load, working memory, and attention for participants navigating a maze via real walking inside a VE. See figure \ref{fig:teaser}.  
    \item Data analysis to investigate relationships among cybersickness, mental load, physical load, working memory, and attention for participants navigating a maze via real walking inside a VE.
    \item Deep learning 
    models trained with VRWalking to classify cybersickness while walking effectively.
    \item SHAP analysis to identify dominant features for classifying cybersickness while walking.
\end{itemize}
\par The deep learning 
models we applied achieved an impressive 95\% accuracy in predicting cybersickness while walking. 
Surprisingly, our analysis did not identify a correlation between physiological factors and the other metrics. However, the SHAP analysis revealed that physiological data remains some of the dominant features in predicting cybersickness, alongside eye tracking.

The rest of the paper is outlined as follows: 2) Background - a description of the related work in our area and the research gaps present in current datasets, 3) Data collection procedure - the details of how the data was collected and the tasks that participants performed, 4) Data Collected - a description of each type of data that was collected,   5) Data Analysis - descriptive statistics, correlations, and analysis of variance within the data, 6) Discussion of the Data Analysis, 7) Case studies in classification - several deep learning models are evaluated for classifying cybersickness, 8) Classification discussion, 9) Limitations and Future work, and 10) Conclusion.
\section{Background}
In this section, we define cybersickness, cognitive load, attention, and working memory, discuss measurement and prediction methods, and highlight the research gaps at the intersection of these areas.
\subsection{Cybersickness}
Cybersickness, a term coined by Stanney et al. \cite{stanney1997cybersickness}, encompasses a range of discomforts experienced by users during virtual environment interactions. Rooted in the sensory conflict theory \cite{laviola2000discussion}, cybersickness arises from inconsistencies between visual and vestibular senses. Studies, such as Mayor et al. \cite{mayor2019comparative}, have indicated reduced cybersickness severity during real walking locomotion in VR, corroborating this theory. Beyond neuro-physiological perspectives, theories like dual-task interference \cite{pashler1994dual} suggest that concurrent task performance exacerbates discomfort. Kasper et al. \cite{kasper2014isolating} noted impaired performance when coordinating tasks in VR. While subjective assessments like the Simulator Sickness Questionnaire (SSQ) \cite{kennedy1993simulator} and Fast Motion Sickness (FMS) scale \cite{keshavarz2011validating} gauge cybersickness severity, physiological and motion data offer objective insights \cite{kim2005characteristic}. Machine learning approaches, as demonstrated by Oh et al. \cite{oh2021machine} and Islam et al. \cite{islam2020automatic}, leverage both subjective and objective data for cybersickness prediction. However, prior research primarily involved stationary or minimally moving users. Our study, in contrast, focuses on continuous real walking locomotion in VR.
\subsection{Cognitive Load}
\par Cognitive Load encompasses the mental and physical effort required to accomplish a task. Cognitive Load Theory (CLT) categorizes cognitive load into subtypes, including intrinsic load or physical load(related to the inherent task difficulty), extraneous load or mental load (involving mental demands), and germane load (related to long-term memory)\cite{sweller2019cognitive}. Much of the research on working memory has primarily centered on reducing extraneous load, as modifying instructional methods seems like a pragmatic approach to alleviating cognitive requirements\cite{schrader2012influence}. Researchers have previously used the Paas Scale to measure mental load\cite{aldekhyl2018cognitive,sweller2018measuring} - a single 9-point Likert scale question rating mental load. But as pointed out by Aldekhyl et al., NASA TLX\cite{hart1988development} is a better measure to understand the subtypes of the cognitive load as defined by the CLT\cite{aldekhyl2018cognitive}. 
\subsection{Working Memory \& Attention}
\par Working memory is a brain mechanism tasked with temporarily storing and manipulating information \cite{baddeley2010herding}. Working memory has limited capacity and can hold only a modest amount of information, whether it be abstract concepts or countable objects\cite{cowan2014working}. To assess working memory, we employed a methodology akin to the digit span test\cite{blackburn1957revised}. 
The connection between cognitive load theory and working memory models is somewhat tenuous and may exhibit inconsistencies\cite{schuler2011role}. Schuler et al. stressed the significance of including working memory as a control variable in research. 
\par Attention is defined as a set of processes that help people comprehend information by giving some elements of the environment (or tasks) priority over others\cite{allport1993attention}. 
The study of human visual attention in VR  can be done using the data that is readily available in the HMD, such as the eye-tracking, head-tracking data, and the stereoscopic video data\cite{upenik2017simple} or through the use of dual-task metrics\cite{ho2005assessing}. However, little research has been done to explore the relationship between attention and the impact of cybersickness on attention while users are walking. Furthermore, contemporary VR researchers have increasingly emphasized the prediction of multimodal visual attention\cite{voinescu2023effectiveness,fathy2023virtual,li2021predicting}. 
However, the availability of labeled data for attention prediction in existing datasets is notably limited.

\subsection{VR datasets}
See table \ref{table:datasets} for an overview of the data types collected in prior open datasets for VR. 
In the VR domain, datasets typically specialize in specific areas like cybersickness or cognitive load, but not typically both. For instance, while SET\cite{islam2022towards} and VR.net\cite{wen2024vr} extensively cover cybersickness, others like VREED\cite{tabbaa2021vreed} and the GW Dataset\cite{kothari2020gaze} touch on cognitive load to varying degrees. Dell et al.\cite{dell2020cognitive} proposed a machine learning approach to predict cognitive workload. Wan et al.\cite{wan2021measuring} used EEG signals to measure cognitive ability. Hadi et al. and Li et al.\cite{nobari2021effect,li2020impact} assess the dual-task performance and multitasking impact on cognitive status, but they only focused on the older adult population. However, there's a clear gap in datasets that comprehensively capture factors such as physical load, mental load, attention, and working memory, offering a holistic view of cognitive status alongside cybersickness.

To bridge this gap, our VRWalking dataset aims to provide detailed labeling, including complex head and eye tracking data, physiological measures, and left-eye VR images. Additionally, pre and post-session Simulator Sickness Questionnaires (SSQs)\cite{kennedy1993simulator} are used to assess cybersickness comprehensively. Moreover, a post-session NASA-TLX questionnaire is employed to evaluate cognitive load. To enhance labeling accuracy, Pass-scale-like\cite{roelofs2004psychometric} questions are also administered during the session, enriching the collected time-stamped data.

Importantly, the aforementioned datasets overlook two critical aspects inherent in real-world VR applications: navigation based on actual walking and multitasking. Many contemporary VR applications, such as firefighter response training, medical simulations, and gaming, rely on real walking for navigation and involve multitasking scenarios. To effectively evaluate VR and ensure its accessibility for all users, it's crucial to acknowledge these factors during data collection.

\section{Data Collection Procedure}
\begin{figure}[h]
    \centering
    \includegraphics[width=\columnwidth]{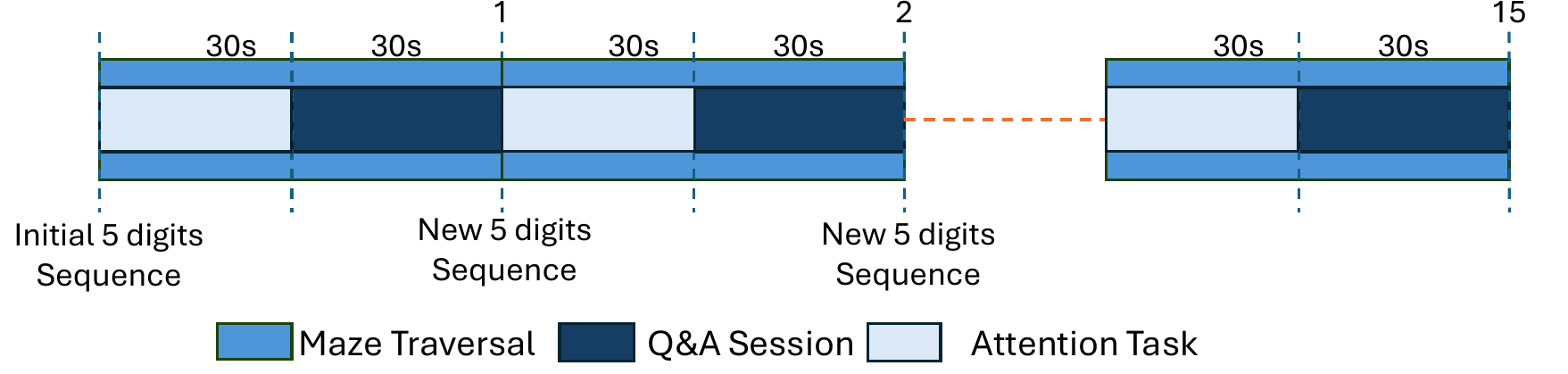}
    \caption{Overview of the Maze Navigation Session}
    \label{fig:overview-maze}
\end{figure}
We collected the data for our VRWalking dataset while participants navigated virtual mazes via real walking for 15 minutes while concurrently performing multiple cognitive tasks. The purpose of this data collection was to investigate the relationships between cybersickness, working memory, attention, and cognitive load. Figure \ref{fig:overview-maze} shows the timeline of the 15-minute session. In the following sections, we provide details of our hypotheses, the virtual environment, the tasks, the study procedure, metrics, and analysis.

\subsection{Participants}
\par We utilized 36 of 39 (23 M, 16 F) participants because the three removed had incomplete sensor data. The participants were recruited from the local area and had a mean age of 25.67 with a standard deviation of 7.22. The participants came from various demographic backgrounds. Twenty-five of 39 participants said they had previously used VR equipment, while thirteen had no prior experience of using VR and one decided not to answer.

\subsection{Hypotheses}
The primary focus of this data collection was to help determine VR's effects on cybersickness, working memory, physical load, and mental load during multiple cognitive tasks and real walking tasks in a VR maze. Our hypotheses were motivated by effects seen in previous literature, which focused primarily on stationary VR. (see background and related work)
    H1: Participants will have a significant rise in cybersickness levels proportionate to their time in VR.
    H2: Working memory will decline, and physical and mental load will increase over time in VR.

\subsection{Virtual Environment}

The virtual environment consisted of a randomly generated maze,  based on a hunt-and-kill algorithm\footnote{https://tinyurl.com/yvr56h9d} as follows: 1) Build a completely walled maze. 2) Start at the beginning of the previous maze's finish. 3) Hunt in a random direction to find a grid location that has not been visited. 4) Break walls between the platforms until the entire maze has been visited. 
The maze contained high stone walls, stone platforms, a grey sky, and no ceiling. The maze was a 4x4 grid that randomly spawned the trophy, the finish condition that transports the user to the next randomly generated maze(Figure~\ref{fig:teaser} middle), each with a minimum navigation distance of at least half of the maze. This ensures that the user has to explore at minimum half of the maze in order to progress to the next level. Thus, all randomly generated mazes have approximately the same difficulty. The virtual maze occupies 3m x 3m of real space, with each platform being about .5m x .5m. 

   
    
\subsection{Study Tasks}
   
The primary task was navigating mazes for 15 minutes. See figure~\ref{fig:teaser} left for an example of a maze. As described in figure \ref{fig:overview-maze}, the whole session was divided into one-minute sections where the maze navigation goes on continuously. In each section, for the first 30 seconds, the participants navigate the maze and complete attention task; in the next 30 seconds, the participants continue navigating the maze, but instead of going through the attention task, they go through a Q\&A session. Each maze contained an average of 4-5 turns, and we instructed the user to stay in motion at all times. This task was to be done for the entire duration of 15 minutes unless the participant decided to quit early. We chose 15 minutes as prior literature has shown that most people will have feelings of cybersickness after 10-15 minutes\cite{melo2018presence}.
  
\subsection{Apparatus}
\par We utilized an HTC-Vive Pro Eye headset to present the virtual maze, offering a display resolution of 1440 x 1600 per eye and operating at a refresh rate of 90Hz. The headset provided a wide field of view, spanning 110 degrees. Audio was delivered through the integrated Vive headphones, and to ensure a seamless VR experience, we configured the HMD with the Vive wireless adapter. The computer used for rendering had the Windows 10 operating system, a 4.35 GHz Intel Core i7 processor, 32 GB DDR3 RAM, and an NVIDIA GeForce RTX 2080 graphics card. The virtual environment was crafted using Unity 3D. To capture eye-tracking data, including gaze direction and pupil diameter, we employed the HTC SRanipal SDK and Tobii HTC Vive Devkit. Additionally, we used the SteamVR lighthouse system to capture head tracking. In order to gather all the data efficiently without interrupting the VR simulation, we used a multi-threaded logging \footnote{https://tinyurl.com/yckknkdw} module within the Unity software\cite{lamb2022eye}. We employed the Neulog\footnote{https://neulog.com/} system to gather heart rate and galvanic skin response data, utilizing two external modular sensors. To ensure convenient data preservation, we employed a Neulog Wi-Fi module for wireless data transmission.
\subsection {Procedure}

\subsubsection{Study Introduction}
Participants read the informed consent form, and we gave them an overview of the different activities during the study. They were also told that they would be able to end the study at any moment if they became too uncomfortable. Upon signing the consent form, participants were then instructed to fill out a background questionnaire, which included an initial SSQ, demographic questions, and disability questions. After fitting participants with the HTC Vive Pro Eye headset, physiological monitoring equipment, and controllers, they were brought into a tutorial that explained the tasks throughout the maze.
    
\subsubsection{Tutorial}
Before the 15-minute VR session, the participants went through a brief 1-minute virtual tutorial explaining the tasks they had to complete. This tutorial included instructions on operating the handheld controller and which button to press during the attention task. The tutorial also ensured that the participants heard the instructions clearly through the HMD's headphones. In the tutorial, the participants were placed in a flat space, and an audio message explained the questions asked during the maze task. After the introductory message was played, the participant was instructed to press the trigger when they heard a keyword to train them for the attention task. They were then instructed to intersect their hand with a 'trophy' (figure \ref{fig:teaser} middle) to complete a maze. 
\subsubsection{Maze Navigation}
 After this training, the participants were then put through an eye calibration to ensure that the system accurately interprets and records eye movements.  The participants were given the target word for the attention metric and the first sequence of five randomly generated digits, as shown in figure \ref{fig:overview-maze}. Once the participants were ready, they were instructed to start navigating the maze. Every 30 seconds, the Q\&A session started, and they answered the three Likert scale questions on cybersickness, mental load, and physical load, recited the previous five digits given, and were given a new five-digit number to remember. The participants navigated the maze via real walking. They also were instructed to complete the attention task throughout the session as they navigated the maze. When at the end of a maze,  they intersected the maze completion marker with their controller and then were teleported to the next maze.
\subsubsection{Post-experience questionnaires}
 After fifteen minutes had passed or the participant decided to quit, they exited the maze and completed another SSQ and a NASA TLX questionnaire. At the end of the session, the participant was paid \$30 per hour and parking fees.
\section{Data Collected}
We collected the following data for our novel VRWalking dataset:
\begin{table*}[h]
\centering
\begin{tabular}{|l|l|l|l|l|l|l|l|l|}
\hline
Dataset            & Exposure Time       & Eye Tracking & Head Tracking & HR           & GSR          & Locomotion   & Cognitive Load & Cybersickness \\ \hline
SET\cite{islam2022towards}                & 7 Minutes           & Yes          & Yes           & Yes           & Yes           & No           & No             & Yes           \\ \hline
\textbf{VRWalking} & \textbf{15 Minutes} & \textbf{Yes} & \textbf{Yes}  & \textbf{Yes} & \textbf{Yes} & \textbf{Yes} & \textbf{Yes}   & \textbf{Yes}  \\ \hline
VREED\cite{tabbaa2021vreed}              & 1-3 Minutes         & Yes          & No            & No           & Yes          & No           & No             & No            \\ \hline
GW Dataset\cite{kothari2020gaze}         & 5 Minutes           & Yes          & Yes           & No           & No           & Yes          & No             & No            \\ \hline
EHTask\cite{9664291}             & 150 Seconds         & Yes          & Yes           & No           & No           & No           & No             & No            \\ \hline
VR.net\cite{wen2024vr}         & 10 Minutes           & Yes            & Yes            & Yes* & No   & Mixed  & No  & Yes
            \\ \hline
\end{tabular}
\caption{Some of the prior open datasets for VR and the data types included. '*' represents partially collected}
\label{table:datasets}
\end{table*}
\par Table \ref{table:datasets} highlights the comprehensive nature of our dataset, encompassing a wide array of features. This dataset provides essential elements for investigating cybersickness, cognitive load, and can be seamlessly integrated into other open datasets to create a more expansive resource for future researchers.

\textit{Attention:} While inside the VR session, as the participants were navigating through the maze, they heard one word at a time every 2-3 seconds randomly generated from a pool of five words  ("Alpha", "Bravo", "Charlie", "Delta", and "Echo"). The participants were given a specific target word at the beginning of the session. Throughout the session, they had to press the instructed button when they heard the words and skip if the word they heard matched the given target word. Two performance metrics were calculated from this task to assess the participant's attention\cite{ho2005assessing}. 
\begin{itemize}
	\item Correct Button Presses: We calculated this as a percentile of how many button presses were correct. The correct button presses included presses on other words and skips on target word. We refer to this as \textit{Attention (Success Rate)}.
	\item Reaction Time: We calculated this as an elapsed time from the word played through the headset and the button presses. This time was then averaged in one-minute windows aligned with the collection of FMS, physical load, and mental load questions. We refer to this as \textit{Attention (Reaction Time)} reported in seconds.
\end{itemize}  
\textit{Working memory load:} Alongside the target word for the attention task, the participants also received a five-digit number played through the headphones. During the Q\&A session, we ask them to repeat the five digits, and at the end of the Q\&A session, we give them a new randomly generated five-digit number. The working memory was calculated as a percentile of the number of digits they repeated correctly. The repeated digit was considered correct if the index and the digit matched the given digit. For example, if the given five-digit number is 57342 and the participant responded 58352, we consider this 5X3X2, where the X represents the missed digits. The performance score will become 60\%. This task is similar to the digit span test\cite{blackburn1957revised} used in previous research\cite{kim2011effect}.

\textit{NASA-TLX:} We took this after the partipants exited the VR session. It consists of six subscales that assess different aspects of workload: mental demand, physical demand, temporal demand, performance, effort, and frustration. Each subscale is rated on a 0-100 scale, with higher scores indicating a higher workload\cite{hart1988development}. 


\textit{Physical Load:} We asked a single question from the NASA-TLX once per minute: How physically demanding is the task on a scale of not demanding/1-very demanding/10?

\textit{Mental Load:} Similar to the Paas scale \cite{sweller2019cognitive}, we asked the following question once per minute: How mentally demanding is the task on a scale of not demanding/1-very demanding/10?

\textit{Fast Motion Sickness Scale (FMS):} We asked the following question once per minute: How sick do you feel on a scale of no cybersickness/1-very intense cybersickness/10?  \cite{keshavarz2011validating,freiwald2020cybersickness}

\textit{SSQ:} We took pre-session and post-session SSQ before and after the VR session respectively. From the 16 questions, we calculated the total score and three subscores: Nausea, Oculomotor, and Disorientation\cite{kennedy1993simulator}.

\textit{Eye Tracking:} The SRanipal SDK(Sensory Reality Affective Neural-integrated Processing SDK) seamlessly incorporates Tobii's eye tracking technology. Through this technology, we gathered detailed eye-tracking data for each participant. 
Importantly, this eye-tracking data was logged along with timestamps and was automatically stopped when the VR session concluded. The eye-tracking data was sampled at a rate of 60 Hz.

\textit{Head Tracking:} For head tracking, we employed the SteamVR Lighthouse tracking system, which consists of base stations emitting laser signals. These signals are picked up by the headset, enabling the system to accurately calculate the headset's position and orientation, thereby providing head-tracking data. This head-tracking data was sampled at a rate of 60 Hz.

\textit{Physiological Readings:} We gathered bio-physiological data (i.e., Heart Rate(HR) and Galvanic Skin Response (GSR)) using external sensors that were affixed to the participant's fingers. To ensure participants had the freedom to move around comfortably, we employed wireless modular sensors provided by Neulog. This data was collected at a 10 Hz sampling rate.

\textit{VR Images:} We recorded point-of-view video footage of participants within the VE they were experiencing. These videos were easily accessible through the HMD and were recorded and saved independently. 

\par Given that head tracking and eye tracking data were sampled at a rate of 60 Hz, while HR and GSR were sampled at 10 Hz, we synchronized their frequencies by downsampling all signals to 1 Hz. This synchronization facilitated coherent analysis across modalities and was necessary for creating the dataset for cybersickness prediction. 

 \section{Data Analysis}
 \subsection{Descriptive Statistics}
\begin{figure}[h]
    \centering
    \includegraphics[width=\columnwidth]{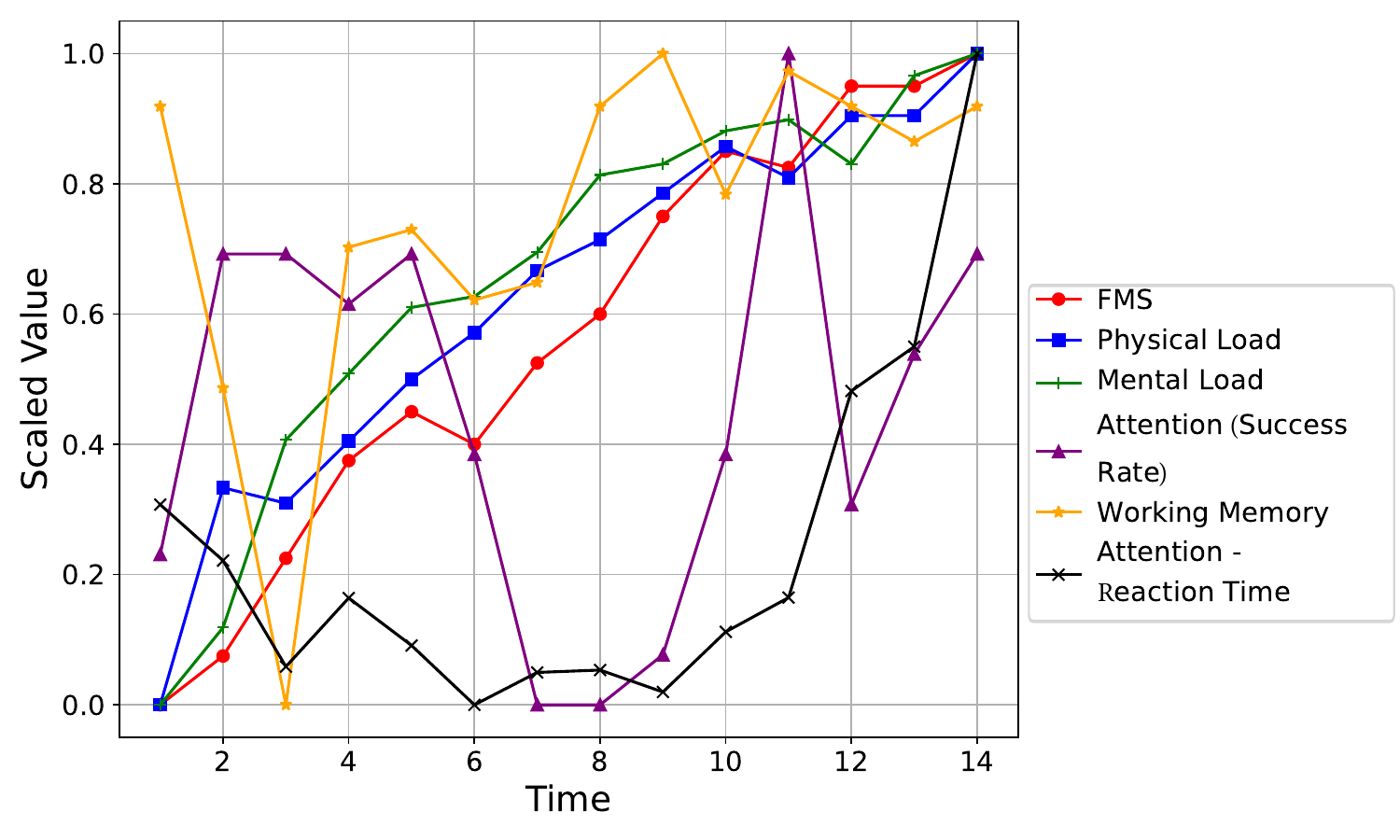}
    \caption{Normalized FMS, Physical Load, Mental Load, Working Memory, Attention (Success Rate), and Attention (Reaction Time) Over Time for one participant}
    \label{fig:fms}
\end{figure}

 Descriptive statistics can be found in Tables \ref{table:desc1} and \ref{table:descriptive-ssq}.  An example of how these data vary over time can be observed in figure \ref{fig:fms}. 
 Based on the descriptive statistics for attention, it appears relatively stable with a low standard deviation. However, Figure \ref{fig:fms} visually demonstrates that while FMS, physical load, mental load, and Attention (Reaction Time) steadily increase, working memory and Attention (Success Rate) exhibit more fluctuations over time. We also noticed high variability in the physiological measurements in general.
\begin{table}[h]
\centering
\caption{Descriptive Statistics}
\label{table:desc1}
\begin{tabular}{|l|l|l|}
\hline
Metric         & Mean  & SD    \\ \hline
FMS            & 1.93 & 1.27    \\ \hline
HR             & 78.13 & 14.25 \\ \hline
GSR            & 3.07  & 2.4   \\ \hline
Physical Load  & 2.11  & 1.24  \\ \hline
Mental Load    & 3.75  & 1.86  \\ \hline
Working Memory & 76.07 & 17.98 \\ \hline
Attention (Success Rate)     & 63.91 & 5.3   \\ \hline
Attention (Reaction Time)  & 0.22 & 0.13 \\ \hline
\end{tabular}%
\end{table}
\par Furthermore, we can deduce from Table \ref{table:desc1} that, on average, the majority of participants experienced a low level of cybersickness as expected for a walking task. Conversely, most participants exhibit elevated levels of mental and physical load compared to cybersickness.
\par To construct Figure \ref{fig:fms} we aggregated the physical load, mental load, and task performance metrics for working memory and attention by computing the average value at each time point across all participants. This means that, for example, the physical load at a given time point represented the average of the physical load values of all 36 participants at that specific time point. Since the value ranges of these metrics differ (e.g., physical load ranges from 1 to 10 while working memory ranges from 0 to 100), we standardized all values to a scale of 0 to 1. This standardization was done to ensure that all metrics could be visually compared on the same scale in the graphs, facilitating a comprehensive overview of the progression of each metric over time (See Figure \ref{fig:fms}).
\begin{center}
\begin{table}[h]
\centering
\caption{Descriptive Statistics of Pre-session and Post-session SSQ}
\label{table:descriptive-ssq}
\begin{tabular}{ |c |c |c| }

\hline
  & Pre-session & Post-session \\
\hline
Nausea         & 0.39$\pm$0.69    & 2.75$\pm$1.92     \\\hline
Oculomotor     & 0.97$\pm$1.32    & 3.94$\pm$3.57    \\\hline
Disorientation & 0.36$\pm$0.68    & 2.33$\pm$2.45     \\\hline
Total Score    & 6.4$\pm$9.32    & 33.76$\pm$27.21   

\\\hline
\end{tabular}
\end{table}
\end{center}
 
\subsection{Differences Over Time}
 A Shapiro-Wilk test on the data indicated that we cannot assume normality. That is why we performed the Wilcoxon signed-rank test to determine the difference between the pre-session and post-session  SSQ subscores and the total scores. All cybersickness subscores indicated a significant increase post-session relative to pre-session: Nausea (V = 16.0, p = $2.02\times10^{-6}$, Cohen’s d = 6.16), Oculomotor (V = 71.0, p = $3.47\times10^{-5}$, Cohen’s d = 15.69), Disorientation (V=18.0, p = $3.41\times10^{-5}$, Cohen’s d = 14.89) and Total Score (V = 57.0, p = $2.60\times10^{-7}$, Cohen’s d = 9.12). The mean scores of each subscore are displayed in Table \ref{table:descriptive-ssq}.
\par We expanded our analysis by stratifying the data based on VR exposure time and participants' FMS scores, conducting Wilcoxon-signed rank tests for paired groups and Mann-Whitney U tests for unpaired groups to assess statistical significance (Table \ref{table:differencesovertime}). In the case of exposure time, we divided the 15-minute sessions into two groups of 7.5 minutes each, comparing FMS and physiological data between them. Results revealed statistical significance between the first and second half of FMS scores, with no significant differences observed in heart rate (HR) and galvanic skin response (GSR). However, when examining the initial and final 30 seconds of HR and GSR, significant differences emerged for GSR, suggesting an impact of exposure time on FMS and GSR.
\subsection{Differences Between Sick and Non-Sick Participants}
Additionally, participants were grouped based on their FMS scores, categorized as 'sick' (FMS score $\geq$ 2) and 'non-sick' (FMS score $<$ 2), with the latter serving as the lower quartile (Q1) reference. Despite observing no significant differences in physiological data between these groups, significant disparities were detected in NASA-TLX's physical demand and annoyance scales (Table \ref{table:sicknotsick}).

\begin{table}[h]
\centering
\caption{Differences in Data over Time from Wilcoxon Signed Rank Tests}
\resizebox{\columnwidth}{!}{%
\begin{tabular}{|l|l|l|}
\hline
Category                                                                                                             & Stats     & Value                                            \\ \hline
\multirow{4}{*}{First Half FMS vs  Second Half FMS}                                                                  & W         & 20                                               \\ \cline{2-3} 
                                                                                                                     & z         & -4.92                                            \\ \cline{2-3} 
                                                                                                                     & p         & 0.0009                                           \\ \cline{2-3} 
                                                                                                                     & Cohen's d & 18.33333333                                      \\ \hline
\multirow{4}{*}{\begin{tabular}[c]{@{}l@{}}First Half Physical Load vs  \\ Second Half Physical Load\end{tabular}}   & W         & 23                                               \\ \cline{2-3} 
                                                                                                                     & z         & -4.87                                            \\ \cline{2-3} 
                                                                                                                     & p         & 0.0001                                           \\ \cline{2-3} 
                                                                                                                     & Cohen's d & 13.41                                            \\ \hline
\multirow{4}{*}{\begin{tabular}[c]{@{}l@{}}First Half Mental Load vs  \\ Second Half Mental Load\end{tabular}}       & W         & 17.5                                             \\ \cline{2-3} 
                                                                                                                     & z         & -4.96                                            \\ \cline{2-3} 
                                                                                                                     & p         & $3.64\times 10^{-6}$ \\ \cline{2-3} 
                                                                                                                     & Cohen's d & 5.75                                             \\ \hline
\multirow{4}{*}{First 30s GSR vs  Second 30s GSR}                                                                    & W         & 93                                               \\ \cline{2-3} 
                                                                                                                     & z         & -3.77                                            \\ \cline{2-3} 
                                                                                                                     & p         & $6.62\times 10^{-5}$ \\ \cline{2-3} 
                                                                                                                     & Cohen's d & 15.5                                             \\ \hline
\end{tabular}%
}\label{table:differencesovertime}
\end{table}
\begin{table}[h]
\centering
\caption{ Mann-Whitney U tests comparing NASA-TLX (physical demand, annoyance), physical load, mental load, Working Memory, and attention between sick participants (13) and non-sick participants(23)}
\resizebox{\columnwidth}{!}{%
\begin{tabular}{|l|l|l|l|l|l|l|}
\hline
Category                                                                        & Mean  & SD    & z-value                & U                      & p-value                 & Cohen's d             \\ \hline
\begin{tabular}[c]{@{}l@{}}Physical Demand-\\ NASA-TLX  (Sick)\end{tabular}     & 3.15  & 1.63  & \multirow{2}{*}{2.50}  & \multirow{2}{*}{225.5} & \multirow{2}{*}{0.009}  & \multirow{2}{*}{0.43} \\ \cline{1-3}
\begin{tabular}[c]{@{}l@{}}Physical Demand-\\ NASA-TLX  (Non-sick)\end{tabular} & 1.87  & 1.01  &                        &                        &                         &                       \\ \hline
Annoyance-NASA-TLX (Sick)                                                       & 5.07  & 2.22  & \multirow{2}{*}{3.80}  & \multirow{2}{*}{265}   & \multirow{2}{*}{0.0001} & \multirow{2}{*}{0.50} \\ \cline{1-3}
\begin{tabular}[c]{@{}l@{}}Annoyance-\\ NASA-TLX (Non-sick)\end{tabular}        & 2.26  & 1.05  &                        &                        &                         &                       \\ \hline
Working Memory (Sick)                                                           & 65.27 & 22.68 & \multirow{2}{*}{-2.10} & \multirow{2}{*}{85.5}  & \multirow{2}{*}{0.03}   & \multirow{2}{*}{0.16} \\ \cline{1-3}
Working Memory (Non-sick)                                                       & 82.17 & 11.22 &                        &                        &                         &                       \\ \hline
\end{tabular}%
}\label{table:sicknotsick}
\end{table}

\section{Data Analysis Discussion}
\par The data analysis section primarily aimed to validate prior hypotheses established across various data collection sessions, thereby demonstrating the comparable potential and applicability of the collected data. We primarily conducted correlation tests to determine the relationship between physical load and mental load encountered during VR sessions, revealing a positive monotonic correlation between them (Coefficient = 0.63). Furthermore, the FMS was validated against post-session SSQ subscores, revealing a positive correlation between FMS and all the subscores. The significance tests were employed to identify features significantly associated with the FMS or other subjective measures gathered during the session. This process provided valuable insights into the features that should be prioritized when collecting similar data in the future. This same motivation drove the subsequent SHAP analysis, which further highlighted the impact of individual features on the target outcomes.
\subsection{Cybersickness}
\par We have observed lower post-session subscores compared to those reported in a previous study\cite{islam2020automatic}. Our hypothesis is that this reduction may be attributed to the continuous movement experienced during the VR session, which could potentially counteract the sensory conflict that often leads to cybersickness. Despite the reduction in cybersickness associated with real walking-based navigation, it's noteworthy that 13 participants (36\%) exhibited an average FMS higher than 2, with 21 participants (58\%) showing an average FMS higher than 1.14 (median).
\par We notice a notable variation in FMS across time, as evidenced by the significance test conducted on the FMS values for the first and second halves. This outcome aligns with the findings from the SSQ assessments conducted before and after the session, where we identified substantial disparities in all subscores as well as the total scores. This collectively suggests that prolonged exposure to virtual reality leads to the onset of cybersickness.
\par We conducted additional analyses to explore the relationships between cybersickness and the collected features. Notably, we identified a robust correlation between the post-session SSQ subscores and the participants' average FMS, affirming the utility of FMS for a detailed understanding of cybersickness. Conversely, we did not uncover significant correlations between HR, GSR and FMS. 
\subsection{Mental Load and Physical Load}
\par Both physical load and mental load showed significant variations when the two groups were categorized based on time. Furthermore, when evaluating the reported physical demand using the NASA TLX, there was a discernible contrast between the non-sick (23 participants) and sick groups (13 participants), with a substantial increase in mean values observed in the latter. However, the verbal reports of physical load and mental load did not exhibit significant variations between the non-sick and sick groups.

We had anticipated differences in both physical load and mental load between the sick and non-sick groups. As expected, we observed the anticipated outcome in the reported physical demand from the NASA-TLX. Both FMS and physical load, as well as mental load, exhibited significant variations over time. However, the absence of differences in physical load and mental load between the sick and non-sick groups suggests that physical and mental load may not be closely related to cybersickness.
\subsection{Working Memory and Attention}
\par Table \ref{table:sicknotsick} highlights a statistically significant difference in working memory between the sick and non-sick groups. However, no significant difference was observed for attention metrics. In Figure \ref{fig:fms}, an upward trend in reaction time over time suggests reduced attention over time. Furthermore, However, we found no indicate no significant relationship between working memory and time. Additionally, it's worth noting a significant correlation between Attention (Success Rate) and Attention (Reaction Time) (p-value $<$ 0.05) was observed during Spearman correlation analysis, with a coefficient of -0.51.
\section{Case Study in Cybersickness Classification}
\par 
To demonstrate potential use cases for our VRWalking dataset, we performed a case study using deep learning models to classify cybersickness. 
\subsection{Categorization}
To support classification, we categorize the FMS  values into three categories based on quantiles. 
\begin{equation}
    Class_t =
\begin{cases}
  \text{Low,} & \text{if } 0 < X \leq Q_1 \\
  \text{Medium,} & \text{if } Q_1 < X \leq Q_2 \\
  \text{High,} & \text{if } X > Q_2 \\
\end{cases}
\label{eq:classes}
\end{equation}
In Equation \ref{eq:classes}, X represents FMS. The quantiles were set as $Q_1=2, Q_2=3$. The deliberate selection of Q1=2 and Q2=3 was made to enhance the classification of cybersickness effectively. Despite the FMS score range spanning from 1 to 10, opting for a smaller range was informed by prior research\cite{islam2020automatic,islam2022towards} and aimed at achieving a more balanced dataset. Typically, FMS levels exceeding 6 are infrequent. Hence, choosing Q1=2 and Q2=3 was ideal for creating a well-balanced classification dataset. 
\subsection{ Neural Network Models: LSTM, GRU \& MLP}
There are three different neural network architectures we used for the cybersickness classification: LSTM (Long Short-Term Memory), GRU (Gated Recurrent Unit), and MLP (Multilayer Perceptron). For sequential data, such as time series data, recurrent neural network (RNN) architectures LSTM and GRU were principally developed. MLP is a feedforward neural network design made up of several interconnected layers of neurons that are also capable of handling time-series data. 
As illustrated in Table \ref{table:nn-architecture}, the LSTM model comprises two LSTM layers for capturing sequential patterns, each followed by a dropout layer for regularization against overfitting. The final dense layer serves multi-class classification. The GRU model incorporates a GRU layer for capturing temporal dependencies, followed by a dropout layer for regularization. It also features an LSTM layer for complex temporal modeling, another dropout layer, and a second GRU layer with an additional dropout. The final layer is a dense layer for classification. The MLP model starts with an initial dense layer for feature extraction, followed by dropout for regularization. It includes more hidden dense layers with a dropout layer between them. The final layer is a dense layer for multi-class classification. 
Our first preference for classification mostly focuses on the use of simple models. This tactical choice not only makes it easier to conduct SHAP Analysis but also makes it easier to interpret. It is important to note that applying SHAP analysis to complicated models like the deep fusion models used for a regression task can be extremely difficult and frequently makes it difficult to identify the dominant features\cite{kundu2022truvr,kundu2023litevr}. Table \ref{table:nn-architecture} illustrates the model architecture we used for the severity classification.
\begin{table}[h]
\centering
\caption{Model Architectures for Classification}
\resizebox{\columnwidth}{!}{%
\begin{tabular}{|cccccc|}
\hline
\multicolumn{6}{|c|}{\textbf{Architecture for the LSTM Model}}                                                                                                                                                      \\ \hline
\multicolumn{1}{|c|}{Layer} & \multicolumn{1}{c|}{Type}    & \multicolumn{1}{c|}{\begin{tabular}[c]{@{}c@{}}Output\\ Shape\end{tabular}} & \multicolumn{1}{c|}{\#Param} & \multicolumn{1}{c|}{Dropout} & Activation \\ \hline
\multicolumn{1}{|c|}{1}     & \multicolumn{1}{c|}{LSTM}    & \multicolumn{1}{c|}{256}                                                    & \multicolumn{1}{c|}{264192}  & \multicolumn{1}{c|}{0.2}     & -          \\ \hline
\multicolumn{1}{|c|}{2}     & \multicolumn{1}{c|}{Dropout} & \multicolumn{1}{c|}{256}                                                    & \multicolumn{1}{c|}{0}       & \multicolumn{1}{c|}{0.2}     & -          \\ \hline
\multicolumn{1}{|c|}{3}     & \multicolumn{1}{c|}{LSTM}    & \multicolumn{1}{c|}{128}                                                    & \multicolumn{1}{c|}{197120}  & \multicolumn{1}{c|}{0.2}     & -          \\ \hline
\multicolumn{1}{|c|}{4}     & \multicolumn{1}{c|}{Dropout} & \multicolumn{1}{c|}{128}                                                    & \multicolumn{1}{c|}{0}       & \multicolumn{1}{c|}{0.2}     & -          \\ \hline
\multicolumn{1}{|c|}{5}     & \multicolumn{1}{c|}{Dense}   & \multicolumn{1}{c|}{3}                                                      & \multicolumn{1}{c|}{387}     & \multicolumn{1}{c|}{-}       & Softmax    \\ \hline
\multicolumn{6}{|c|}{Total No. of Params: 461,699}                                                                                                                                                                  \\ \hline
\multicolumn{6}{|c|}{\textbf{Architecture for the GRU Model}}                                                                                                                                                            \\ \hline
\multicolumn{1}{|c|}{1}     & \multicolumn{1}{c|}{GRU}     & \multicolumn{1}{c|}{32}                                                     & \multicolumn{1}{c|}{3360}    & \multicolumn{1}{c|}{0.2}     & -          \\ \hline
\multicolumn{1}{|c|}{2}     & \multicolumn{1}{c|}{Dropout} & \multicolumn{1}{c|}{32}                                                     & \multicolumn{1}{c|}{0}       & \multicolumn{1}{c|}{0.2}     & -          \\ \hline
\multicolumn{1}{|c|}{3}     & \multicolumn{1}{c|}{LSTM}    & \multicolumn{1}{c|}{64}                                                     & \multicolumn{1}{c|}{18816}   & \multicolumn{1}{c|}{0.2}     & -          \\ \hline
\multicolumn{1}{|c|}{4}     & \multicolumn{1}{c|}{Dropout} & \multicolumn{1}{c|}{64}                                                     & \multicolumn{1}{c|}{0}       & \multicolumn{1}{c|}{0.2}     & -          \\ \hline
\multicolumn{1}{|c|}{5}     & \multicolumn{1}{c|}{GRU}     & \multicolumn{1}{c|}{128}                                                    & \multicolumn{1}{c|}{74496}   & \multicolumn{1}{c|}{0.2}     & -          \\ \hline
\multicolumn{1}{|c|}{6}     & \multicolumn{1}{c|}{Dropout} & \multicolumn{1}{c|}{128}                                                    & \multicolumn{1}{c|}{0}       & \multicolumn{1}{c|}{0.2}     & -          \\ \hline
\multicolumn{1}{|c|}{7}     & \multicolumn{1}{c|}{Dense}   & \multicolumn{1}{c|}{3}                                                      & \multicolumn{1}{c|}{387}     & \multicolumn{1}{c|}{-}       & Softmax    \\ \hline
\multicolumn{6}{|c|}{Total No. of Params: 97,059}                                                                                                                                                                   \\ \hline
\multicolumn{6}{|c|}{\textbf{Architecture for the MLP Model}}                                                                                                                                                            \\ \hline
\multicolumn{1}{|c|}{1}     & \multicolumn{1}{c|}{Dense}   & \multicolumn{1}{c|}{32}                                                     & \multicolumn{1}{c|}{2368}    & \multicolumn{1}{c|}{0.25}    & ReLU       \\ \hline
\multicolumn{1}{|c|}{2}     & \multicolumn{1}{c|}{Dropout} & \multicolumn{1}{c|}{32}                                                     & \multicolumn{1}{c|}{0}       & \multicolumn{1}{c|}{0.25}    & -          \\ \hline
\multicolumn{1}{|c|}{3}     & \multicolumn{1}{c|}{Dense}   & \multicolumn{1}{c|}{128}                                                    & \multicolumn{1}{c|}{4224}    & \multicolumn{1}{c|}{0.25}    & ReLU       \\ \hline
\multicolumn{1}{|c|}{4}     & \multicolumn{1}{c|}{Dropout} & \multicolumn{1}{c|}{128}                                                    & \multicolumn{1}{c|}{0}       & \multicolumn{1}{c|}{0.25}    & -          \\ \hline
\multicolumn{1}{|c|}{5}     & \multicolumn{1}{c|}{Dense}   & \multicolumn{1}{c|}{32}                                                     & \multicolumn{1}{c|}{4128}    & \multicolumn{1}{c|}{0.25}    & ReLU       \\ \hline
\multicolumn{1}{|c|}{6}     & \multicolumn{1}{c|}{Dropout} & \multicolumn{1}{c|}{32}                                                     & \multicolumn{1}{c|}{0}       & \multicolumn{1}{c|}{0.25}    & -          \\ \hline
\multicolumn{1}{|c|}{7}     & \multicolumn{1}{c|}{Dense}   & \multicolumn{1}{c|}{8}                                                      & \multicolumn{1}{c|}{264}     & \multicolumn{1}{c|}{0.25}    & ReLU       \\ \hline
\multicolumn{1}{|c|}{8}     & \multicolumn{1}{c|}{Dropout} & \multicolumn{1}{c|}{8}                                                      & \multicolumn{1}{c|}{0}       & \multicolumn{1}{c|}{0.25}    & -          \\ \hline
\multicolumn{1}{|c|}{9}     & \multicolumn{1}{c|}{Dense}   & \multicolumn{1}{c|}{3}                                                      & \multicolumn{1}{c|}{27}      & \multicolumn{1}{c|}{-}       & Softmax    \\ \hline
\multicolumn{6}{|c|}{Total No. of Params: 11,011}                                                                                                                                                                   \\ \hline
\end{tabular}%
}
\label{table:nn-architecture}
\end{table}
\subsection{Training Setup}
\par We employed a 10-fold cross-validation technique to train and assess the performance of our proposed model, following a methodology similar to previous studies \cite{bengio2003no, maji2011action,padmanaban2018towards}. In the k-fold cross-validation process, the dataset is divided into k subsets. One of these subsets is used for testing the model while the remaining (k-1) subsets serve as the training data \cite{james2013introduction}. This procedure is repeated k times, each time using a different subset as the test data and the rest as the training data. This approach helps mitigate any potential dataset bias \cite{bengio2003no}.
\par In our data preprocessing approach, we applied data normalization to the eye tracking data, ensuring consistent scaling across all observations. Additionally, we employed exponential smoothing techniques to denoise the HR and GSR signals, drawing inspiration from prior research methodologies focused on cybersickness prediction\cite{islam2022towards,islam2020automatic,islam2021cybersickness}.
\par To optimize the model parameters during training, we allocated 30\% of the training dataset as validation data for each fold \cite{ripley2007pattern}. We employed the Adam optimizer with a training duration of 200 epochs and a batch size of 256, all configured through hyperparameter tuning. We used categorical cross-entropy as the loss function. To safeguard against overfitting, we implemented an early-stopping mechanism with a patience setting of 30 during the model training \cite{caruana2000overfitting}.
\subsection{Classification Results}


\begin{table*}[h]
\centering
\caption{Mean Accuracy, Precision and Recall of the 10-Fold Cross Validation on Cybersickness Classification for Different Models}
\label{table:model-performance}
\begin{tabular}{|l|l|l|lll|lll|}
\hline
\multicolumn{1}{|c|}{\multirow{2}{*}{Type}} & \multirow{2}{*}{Used Models} & \multirow{2}{*}{\% Accuracy} & \multicolumn{3}{c|}{\% Precision}                              & \multicolumn{3}{c|}{\% Recall}                                 \\ \cline{4-9} 
\multicolumn{1}{|c|}{}                               &                              &                              & \multicolumn{1}{l|}{Low}  & \multicolumn{1}{l|}{Medium} & High & \multicolumn{1}{l|}{Low}  & \multicolumn{1}{l|}{Medium} & High \\ \hline
\multirow{4}{*}{$CS_t$}                              & GRU                          & 95                           & \multicolumn{1}{l|}{97}   & \multicolumn{1}{l|}{92}     & 95   & \multicolumn{1}{l|}{98}   & \multicolumn{1}{l|}{88}     & 93   \\ \cline{2-9} 
                                                     & LSTM                         & 95                           & \multicolumn{1}{l|}{96}   & \multicolumn{1}{l|}{90}     & 97   & \multicolumn{1}{l|}{98}   & \multicolumn{1}{l|}{89}     & 90   \\ \cline{2-9} 
                                                     & MLP                          & 82                           & \multicolumn{1}{l|}{98}   & \multicolumn{1}{l|}{62}     & 69   & \multicolumn{1}{l|}{83}   & \multicolumn{1}{l|}{72}     & 97   \\ \cline{2-9} 
                                                     \hline
\end{tabular}
\label{table:modelperformance}
\end{table*}

The mean accuracy, precision, and recall for 10-fold cross-validation are listed in Table \ref{table:model-performance}. During the classification task, GRU and LSTM models achieved the highest accuracy. For the mental load classification, LSTM outperformed the GRU model, achieving an accuracy of 95\%.
\subsection{SHAP Analysis}
\begin{figure}
\centering
\includegraphics[width=0.8\linewidth]{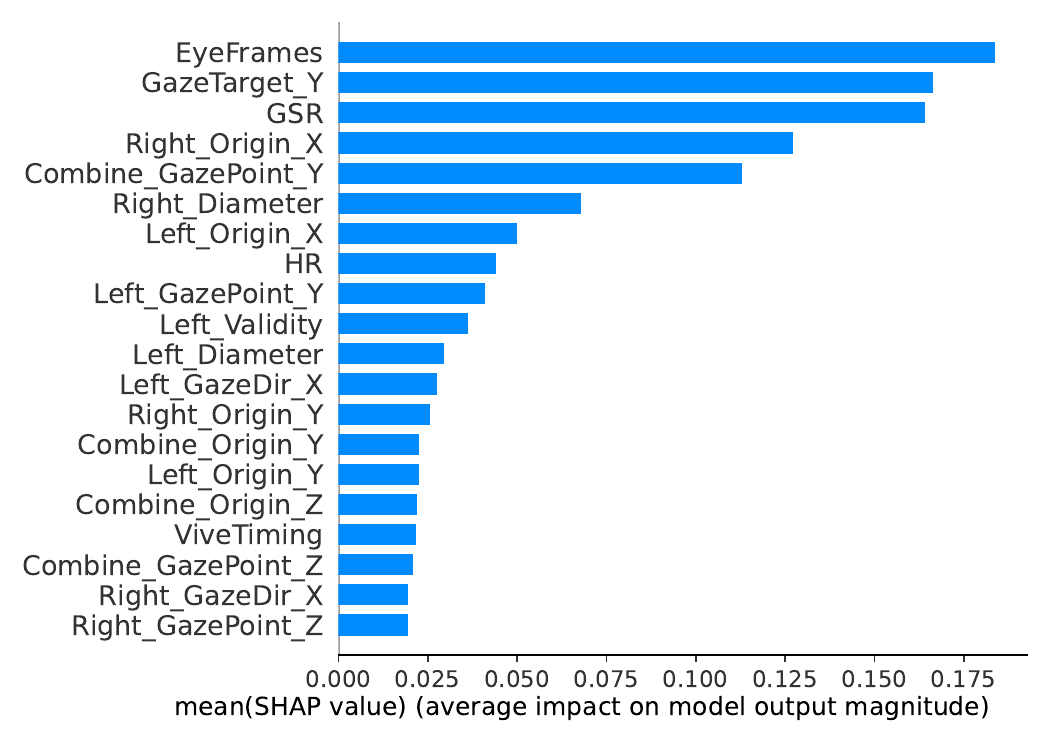}
\caption{Cybersickness Classification (GRU)}
\label{fig:SHAP-CS-GRU}
\end{figure}

In order to further analyze and better understand the performance of the models, we employed SHapley Additive Explanations (SHAP). SHAP is a post-hoc explanation method that ranks the input features. The goal of SHAP is to explain the prediction of a given DL model (e.g., LSTM, GRU, MLP, etc.) for a given set of input samples (e.g., eye and head tracking, HR, etc.). The model architecture for the used DL models is summarized in Table \ref{table:nn-architecture}. Mangalathu et al. emphasize that while complex deep learning models can achieve high accuracy, interpreting them through SHAP analysis poses challenges\cite{mangalathu2020failure}. These models often possess intricate internal representations and intricate feature interactions, rendering the provision of clear and easily understandable explanations for model predictions using SHAP a daunting task. That is why we have used some simpler DL models in our classification, and the global explanation done by the SHAP analysis shows us the dominant features for cybersickness classification. We have listed the DL models that achieved the best performance during the classification task in table \ref{table:modelperformance}.  
Although during the correlation test, we did not find any significant correlation between HR, GSR, and FMS, from the SHAP analysis, we can see that both HR and GSR were dominant features when predicting Cybersickness Figure  \ref{fig:SHAP-CS-GRU} illustrates the dominant features.  


\section{Cybersickness Classification Discussion}
\par The LSTM, GRU, and MLP models demonstrated outstanding predictive performance for all attributes, achieving exceptional accuracy. Notably, the MLP had a lower performance compared to the LSTM and GRU models, but it's important to highlight that the GRU had 8.8 times more parameters, and the LSTM had 42 times more parameters than the MLP. This led to significantly reduced training times for the MLP.



\par When employing the basic LSTM, GRU, and MLP models, we observed a cybersickness severity classification on par with state-of-the-art results, such as Kundu et al.'s \cite{kundu2023litevr} achievement of 94\% accuracy with an LSTM. Despite the data collection occurring in a novel environment with navigation and cognitive tasks and the absence of initial resting baseline measurements for data normalization, our deep learning models trained with our dataset delivered an outstanding performance, achieving an impressive accuracy rate of 95\%. The LSTM and GRU models we used exhibited state-of-the-art level precision and recall for the individual classes compared to prior research\cite{kundu2023litevr}. We anticipate that incorporating a resting baseline and employing more sophisticated models could yield even better outcomes, especially considering that previous research demonstrated the superiority of deep fusion models over basic deep learning models\cite{islam2021cybersickness}. We initially anticipated that the swift movements associated with the navigation task might disrupt the collection of physiological data. Surprisingly, after employing exponential smoothing to standardize the readings, they emerged as some of the most influential features in predicting cybersickness. Additionally, we speculated that the navigational task compelled participants to engage in frequent head movements, introducing considerable variability in the head-tracking data. This may explain why head tracking was not as effective in predicting cybersickness as observed in prior research\cite{islam2022towards,kundu2023vr}.
\section{Limitations and Future Work}
\par 
Our dataset was not balanced with respect to gender, which is known to correlate with cybersickness\cite{macarthur2021you,melo2021impact} and potentially other data types that we collected. To address this, we intend to collect data from a more diverse and gender-balanced participant group to improve generalizability. Furthermore, the attention metric we utilized is limited in that it does not provide insights into sustaining visual attention\cite{rosenberg2013sustaining}. Therefore, we aim to incorporate metrics that assess reaction time from eye tracking to gain a better understanding of sustaining visual attention.
\par Additionally, we exclusively utilized basic deep learning models in this study. However, we anticipate achieving improved results by employing advanced models, such as deep fusion models, which may be better equipped to learn from time series data\cite{islam2021cybersickness}. Moreover, we only used a maze virtual environment with real walking in the data collection. We plan to evaluate different virtual environments with different locomotion interfaces in later studies.

In the future, our aim is to extend the use of these models to classify cognitive effects on working memory and attention and for precise value prediction through regression. Moreover, in line with previous research, we plan to explore the application of SHAP for model reduction. This would enable us to develop efficient and lightweight models suitable for deployment on VR headsets. This approach would empower us to effectively predict both cybersickness and cognitive load without the need for external hardware. Moreover, we plan on doing regression instead of classification as that is better suited for cybersickness prediction and  and for the classification task we plan on conducting statistical analysis to calculate the categorization boundary for a better generalizable approach. 
\section{Conclusion}
\par In conclusion, we collected a dataset to understand the relationship between cybersickness, attention, mental load, physical load, and working memory in VR. Participants were tasked with maze navigation via real walking for 15 minutes. Throughout the experiment, we collected head-tracking, eye-tracking, HR, GSR, and self-reported data on cybersickness,  mental load, and physical load while users performed working memory and attention tasks. Results suggested that participants grew more sick over time. Moreover, the participants who were sicker experienced a significantly higher demand on their working memory performance and their self-reported physical load than participants who were less sick. Next, as an example of the use of the dataset, we developed several deep learning models with the intention of predicting cybersickness based on head tracking, eye tracking, GSR, and HR sensor data as input. Using several simple deep learning models, GRU, LSTM, and MLP, we were able to effectively classify cybersickness with 95\% accuracy. Lastly, we performed a SHAP analysis to identify which features impacted our classifiers the most. Ultimately, VR developers could utilize our dataset to create virtual environments and predictive models that can effectively adapt to the user's current state of mind, which could lead to improved and more personalized VR applications.
\acknowledgments{
 This work was supported in part by Army Research Lab awards W911NF2310401 and W911NF2420035, and National Science Foundation awards 2211785 and 2316240.}

\bibliographystyle{abbrv-doi}

\bibliography{template}
\end{document}